\documentclass[onecolumn,showpacs,preprintnumbers,amsmath,amssymb]{revtex4}
\usepackage{graphicx}
\usepackage{color}
\usepackage{bm}
\usepackage{slashed}

\begin{document}

\title{Fermion Bag Approach for Massive Thirring Model at Finite Density}

\author{Daming Li}
\email{lidaming@sjtu.edu.cn}

\affiliation{School of Mathematical Sciences, Shanghai Jiao Tong
University, Shanghai, 200240, China}

\date{\today}

\begin{abstract}
We consider the 2+1 dimensional massive Thirring model with one
flavor at finite density. Two numerical methods, fermion bag
approach and complex Langevin dynamics, are used to calculate the
chiral condensate and fermion density of this model. The numerical
results obtained by fermion bag approach are compared with those
obtained by complex Langevin dynamics. They are also compared with
those obtained under phase quenched approximation. We show that in
some range of fermion coupling strength and chemical potential the
sign problem in fermion bag approach is mild, while it becomes
severe for the complex Langevin dynamics.
\end{abstract}

\pacs{05.50.+q, 71.10.Fd,02.70.Ss}

\keywords{Thirring model, finite density, complex Langevin dynamics,
fermion bag approach}

\maketitle

\section{Introduction}
The sign problem remains one of the biggest challenges in many
fields, e.g., polymer field theory in condensed matter physics
\cite{Fredrickson_2006}, lattice field theory in high energy
physics. The usual sampling methods, e.g., Langevin dynamics and
Monte Carlo method, fail for the sign problem due to the high
oscillation of complex action, where the Boltzmann factor can not be
regarded as the probability density. Because of the introduction of
fields necessary to decouple repulsive interaction between monomer, the sign problem
can not be avoided for polymer field theory \cite{Daming_346}. For the lattice
field theory in high energy physics,  three reasons will always lead
to the complex action: (1) grand partition function with finite
density; (2) fermion systems; (3) topological terms in the action.

To overcome the sign problem, the complex Langevin (CL) dynamics,
which is obtained from the complexification of the Langevin
dynamics, was used. The CL is rather successful in XY model
\cite{Aarts_2010_0617}, Bose gas \cite{Aarts_2009_2089}, Thirring
model \cite{Pawlowski_2013_2249}, Abelian and
 Non-Abelian lattice gauge model \cite{Flower_1986},
QCD model \cite{Sexty_2014_7748}, and its simplified model including
one link U(1) model, one link SU(3) model, QCD model in the heavy
mass limit \cite{Aarts_2008_1597}, one link SU(N) model
\cite{Aarts_1212.5231}, SU(3) spin model \cite{Aarts_2012_4655},
Polykov chain model \cite{Aarts_2013_6425}. It was also applied to
quantum fields in nonequilibrium \cite{Berges_0508030} and in real
time \cite{Berges_0609058}\cite{Berges_0708.0779}. For some range of
chemical potential and large fluctuation, the complex Langevin may
fail, e.g., the XY model at finite chemical potential for large
fluctuation) \cite{Aarts_2010_3468} and in the Thirring model in 0+1
dimension \cite{Pawlowski_2013_094503}. Unfortunately from early
studies of complex Langevin evolutions
\cite{Hamber_330}\cite{Flower_330}\cite{Ilgenfritz_327} until this
day, the convergence properties of complex Langevin equations are
not well understood. Recently Aarts etc. provided a criterion for
checking the correctness of the complex Langevin dynamics
\cite{Aarts_2011_3270}.
The recent discussion about complex Langevin dynamics can be found in Ref.
\cite{Pehlevan_0710.3756}\cite{Pehlevan_0902.1503}
\cite{Pehlevan_0710.1256}
\cite{Aarts_0912.3360}\cite{Aarts_1306.3075}\cite{Mollgaard_1309.4335}
\cite{Aarts_1407.2090}\cite{Nishimura_1504.08359}\cite{Tsutsui_1508.04231}\cite{Hayata_1511.02437}.

Since the partition function is always real, it is possible to find
suitable variables to represent this partition function with real
action. This is called the dual variable method. It is successfully
applied to many models, including Bose gas \cite{Gattringer_2013},
SU(3) spin model \cite{Mercado_2012}, U(1) and Z(3) gauge Higgs
lattice theory \cite{Mercado_2013}, massive lattice Schwinger model
\cite{Gattringer_2015}, O(3), O(N) and CP(N-1) model
\cite{Bruckmann_05482}\cite{Bruckmann_2015}\cite{Wolff_0908.0284}\cite{Wolff_1001.2231},
fermion bag approach \cite{Chandrasekharan_1304.4900},
 4-fermion lattice theory, including massless Thirring model
  \cite{Chandrasekharan_2011_5276}, Gross-Neveu model
  \cite{Chandrasekharan_2012_6572}, Yukawa model
  \cite{Chandrasekharan_1205.0084},
  Non-Abelian Yang-Mills model\cite{Oeckl_0008095}\cite{Cherrington_0705.2629}, and its coupling with fermion field
  \cite{Cherrington_0710.0323},
lattice chiral model, and Sigma model \cite{Pfeiffer_033501}.
 For the recent progress of solving the sign problem for the nonrelativistic fermion systems, see Ref.
\cite{Li_241117}\cite{Li_1601.05780}\cite{wei_250601}\cite{Banerjee_2014}\cite{Cohen_2015}\cite{Huffman_2014}\cite{Huffman_2016}.

For the fermion systems, the dual method is called fermion bag
approach \cite{Chandrasekharan_1304.4900}. This numerical method not
only overcome the sign problem for model with small chemical
potential, but also a high computational efficiency is achieved for
the small or large interaction between fermions. We study the 2+1
dimensional massive Thirring model at finite density, cf.
\cite{Spielmann_2010}, which can be regards as the effective
theories of high temperature superconductors and graphene, see e.g.,
references given in \cite{Gies_1006.3747}. We have studied this
model at finite density in 0+1 dimension and compared the complex
Langevin dynamics and fermion bag approach \cite{Daming_0710.0323}.
 In this paper we continue to compare the complex
Langvin dynamics and the fermion bag approach for the massive
Thirring model at finite density in 2+1 dimension.

The arrangement of the paper is as follows. In section \ref{model},
the Fermion bag approach for Thirring model is presented and the
chiral condensate and fermion density are obtained. In section
\ref{Langevin dynamics}, the complex Langevin dynamics is given for
this model by introducing bosonic variable. In section
\ref{results}, the chiral condensate and fermion density are
calculated by these two methods and are compared with each other.
Conclusions are given in section \ref{conclusion}.

\section{Thirring model}\label{model}
The lattice partition function for the massive Thirring model at the
finite density in $d$ dimensional lattice $\Lambda =
\{x=(x_0,\cdots,x_{d-1}), x_i=0,\cdots,N-1, i=0,\cdots,d-1\} $ with
even $N$ reads
\begin{eqnarray}\label{2016_5_7_0}
 Z = \int d\bar\psi d\psi e^{-S}
\end{eqnarray}
where $d\bar\psi d\psi = \prod_{x\in \Lambda} d\bar\psi(x) d\psi(x)$
is the measure of the Grassmann fields $\psi = \{\psi(x)\}_{x\in
\Lambda}$ and $\bar\psi=\{\bar\psi(x)\}_{x\in \Lambda}$. We adopt
anti-periodic condition for $\psi$ and $\bar\psi$ in $x_0$ direction
and periodic condition in the other directions
\begin{eqnarray}\label{2016_5_7_0_0}
\psi(x+kN\hat\alpha) = (-1)^{k\delta_{\alpha,0}}\psi(x),
\bar\psi(x+kN\hat\alpha) = (-1)^{k\delta_{\alpha,0}}\bar\psi(x),
x\in \Lambda, k = 0, \pm 1, \cdots,
\end{eqnarray}
where $\hat\alpha$ denotes the unit vector in $\alpha$ direction.
 The action $S$ in
(\ref{2016_5_7_0}) is
\begin{eqnarray}\label{2016_5_7_1}
 S = \sum_{x,y\in\Lambda} \bar\psi(x)D_{x,y}\psi(y) - U \sum_{x\in\Lambda,\alpha=0,\cdots,d-1}
\bar\psi(x)\psi(x)\bar\psi(x+\hat\alpha)\psi(x+\hat\alpha)
\end{eqnarray}
with nonnegative coupling constant $U$ between fermions. The fermion
matrix $D$, which depends on the fermion mass $m$ and chemical
potential $\mu$, is given by
\begin{eqnarray}\label{2016_5_7_2}
D_{x,y}&=& D(\mu,m)_{x,y}  \nonumber  \\ &   = &
\sum_{\alpha=0,\cdots,d-1} \frac{\eta_{x,\alpha}}{2}( e^{\mu
\delta_{\alpha, 0}} s^{+}_{x,\alpha} \delta_{x+\hat\alpha,y}-e^{-\mu
\delta_{\alpha, 0}}s^{-}_{x,\alpha}\delta_{x,y+\hat\alpha})+
m\delta_{x,y}\\ &=& \left\{
  \begin{array}{l l}
\frac{\eta_{x,\alpha}}{2}e^{\mu
\delta_{\alpha, 0}}s^{+}_{x,\alpha}  & \quad \text{if \ $y=x+\hat\alpha$ }\\
-\frac{\eta_{x,\alpha}}{2}e^{-\mu
\delta_{\alpha, 0}}s^{-}_{x,\alpha}  & \quad \text{if \ $y=x-\hat\alpha$ }\\
m  & \quad \text{if \ $y=x$ }\\
0 &  \quad \text{Otherwise}\\
   \end{array} \right. \nonumber
\end{eqnarray}
with staggered phase factor $\eta_{x,0}=1$,
$\eta_{x,\alpha}=(-1)^{x_0+\cdots+x_{\alpha-1}}$, $\alpha=1,\cdots,
d-1$, satisfying $\eta_{x+\hat\alpha,\alpha}=\eta_{x,\alpha}$.  The
boundary condition for $\psi$ and $\bar \psi$ are accounted for by
the sign functions $s^+$ and $s^-$
\begin{eqnarray}
 s^{+}_{x,\alpha} =  \left\{
  \begin{array}{l l}
-1  &  \text{if \ $\alpha=0$ \ and $x_0=N-1$}\\
1 &   \text{Otherwise}\\
   \end{array} \right.,   s^{-}_{x,\alpha} =  \left\{
  \begin{array}{l l}
-1  &  \text{if \ $\alpha=0$ \ and $x_0=0$}\\
1 &   \text{Otherwise}\\
   \end{array} \right.
\end{eqnarray}
with periodic extension for $s^{+}$ and $s^{-}$ with respect to $x$ for
any direction $\alpha$. These two sign functions satisfies $ s^{+}_{x, \alpha}
=  s^{-}_{x+\hat\alpha, \alpha}$ for any lattice $x$ and any direction
$\alpha$.  

The fermion matrix has two kind of symmetries with respect to $\mu$
and $m$ which leads to the symmetry of the determinant $\det D$
(Note that $N$ is even)
\begin{eqnarray}\label{2016_5_7_4}
D(\mu,m)_{x,y}=-D(-\mu,-m)_{y,x} \Longrightarrow \det D(\mu,m) =
\det D(-\mu,-m)
\end{eqnarray}
\begin{eqnarray}\label{2016_5_7_5}
 \varepsilon_x   D(\mu,m)_{x,y} \varepsilon_y  = -  D(\mu,-m)_{x,y} \Longrightarrow \det
D(\mu,m) = \det D(\mu,-m)
\end{eqnarray}
where $ \varepsilon_x=(-1)^{x_0+\cdots+x_{d-1}}$ is the parity of
site $x$. Thus it is sufficient to study the massive Thirring model
for $\mu\geq 0$ and $m\geq 0$.

The fermion bag approach for the Thirring model is based on the high
temperature expansion of the interacting term
\begin{eqnarray}\label{2016_5_7_6}
&&\exp\Big(U \sum_{x\in\Lambda,\alpha=0,\cdots,d-1}
\bar\psi(x)\psi(x)\bar\psi(x+\hat\alpha)\psi(x+\hat\alpha)\Big)\nonumber\\
& = &  \prod_{x\in\Lambda,\alpha=0,\cdots,d-1}\sum_{k_{x,
\alpha}=0}^1(U\bar\psi(x)\psi(x)\bar\psi(x+\hat \alpha)\psi(x+\hat
\alpha))^{k_{x, \alpha}}
\end{eqnarray}
Inserting this expansion into the partition function in
(\ref{2016_5_7_0}), one has an expansion of $Z$ with respect to $U$
\begin{eqnarray}\label{2016_5_7_7}
 Z = \sum_{k=(k_{x,\alpha})}   U^{j} C(x_1,\cdots,x_{2j})
\end{eqnarray}
where the summation is taken over all configuration $k$ with
$k_{x,\alpha}=0,1$ for all two neighboring sites $(x,x+\hat\alpha)$
and $ \sum_{\alpha=0}^{d-1}k_{x,\alpha}$ must be 0 or 1 for all site
$x$. If $k_{x,\alpha}=1$, we say there is a bond connecting $x$ and
$x+\hat\alpha$; otherwise, there are no bonds connecting them.  For
a given configuration $k$, for example, there are $j$ bonds
$(x_1,x_2),\cdots,(x_{2j-1},x_{2j})$ connecting $2j$ different
sites, and the weight in (\ref{2016_5_7_7}) depending on these $2j$
different sites $\{x_i\}_{i=1}^{2j}$ is
\begin{eqnarray}\label{2016_5_7_8}
C(x_1,\cdots,x_{2j}) &=& \int d\bar\psi d\psi \exp\Big(-\sum_{x,y}
\bar\psi(x)D_{x,y}\psi(y)\Big) \bar\psi(x_1)\psi(x_1)\cdots
\bar\psi(x_{2j})\psi(x_{2j}) \nonumber  \\  & = &  \det D\det
G(\{x_1,\cdots,x_{2j}\})  =  \det D(\backslash
\{x_1,\cdots,x_{2j}\})
\end{eqnarray}
where $G(\{x_1,\cdots,x_{2j}\})$ is a $(2j)\times (2j)$ matrix of
propagators between $2j$ sites $x_i$, $i=1,\cdots,2j$, whose matrix
element are $G(\{x_1,\cdots,x_{2j}\})_{i,l}=D^{-1}_{{x_{i}},x_{l}}$,
$i,l=1,\cdots,2j$. The matrix $G(\{x_1,\cdots,x_{2j}\})$ depends on
the order of $\{x_1,\cdots,x_{2j}\}$, but it's determinant does not.
$D(\backslash \{x_1,\cdots,x_{2j}\})$ is the $(N^d-2j)\times
(N^d-2j)$ matrix which is obtained by deleting rows and columns
corresponding to sites $x_1, \cdots, x_{2j}$. The first equality in
(\ref{2016_5_7_8}) holds due to the basic Gaussian integration for
the Grassmann variables \cite{Rothe_2005}. In the second equality of
(\ref{2016_5_7_8}) we expand the exponential and then integrating
the Grassmann variables $\{x_i\}_{i=1}^{2j}$. The average number of
bonds depends on the interaction strength $U$ between fermions. If
$U$ is small, there are few bonds between two neighboring sites, we
use $G(\{x_1,\cdots,x_{2j}\})$ to calculate $C(x_1,\cdots,x_{2j})$;
Otherwise, $U$ is large and there are many occupied bonds between
neighboring sites and thus $D(\backslash \{x_1,\cdots,x_{2j}\})$ is
used to calculate $C(x_1,\cdots,x_{2j})$.      For any number of
different sites $\{x_i\}_{i=1}^n$, the function
$C=C(x_1,\cdots,x_n;D(\mu,m))$, depends on the fermion matrix
$D(\mu,m)$.

Because of the symmetry (\ref{2016_5_7_4}) and (\ref{2016_5_7_5}) of
$D$, the function $C$ for any different sites $\{x_i\}_{i=1}^n$ have
the symmetry (APPENDIX \ref{Appendix_1})
\begin{eqnarray}\label{2016_5_8_0}
&&  C(x_1,\cdots,x_{n}; D(\mu,m)) =
 (-1)^n C(x_1,\cdots,x_{n}; D(\mu,-m)) \nonumber \\ &  =  &  C(x_1,\cdots,x_{n};
  D(-\mu,m)) = (-1)^n C(x_1,\cdots,x_{n};
  D(-\mu,-m))
\end{eqnarray}
for any real number $\mu$ and $m$. According to the representation
of the partition function in (\ref{2016_5_7_7}), where $n=2j$ is
even, the weight $C$ becomes nonnegative for any $\mu$ and $m$ if $
C(x_1,\cdots,x_{n}; D(\mu,m))$ is nonnegative for any even number of
sites $(x_1,\cdots,x_{n})$ and for any nonnegative $\mu$ and
nonnegative $m$. Unfortunately $C$ is not always positive and thus
the sign problem still exist. But we want to justify that the sign
problem in the representation of (\ref{2016_5_7_7}) is rather mild.

\begin{figure}
\centering
\includegraphics[width=8cm,height=6cm]{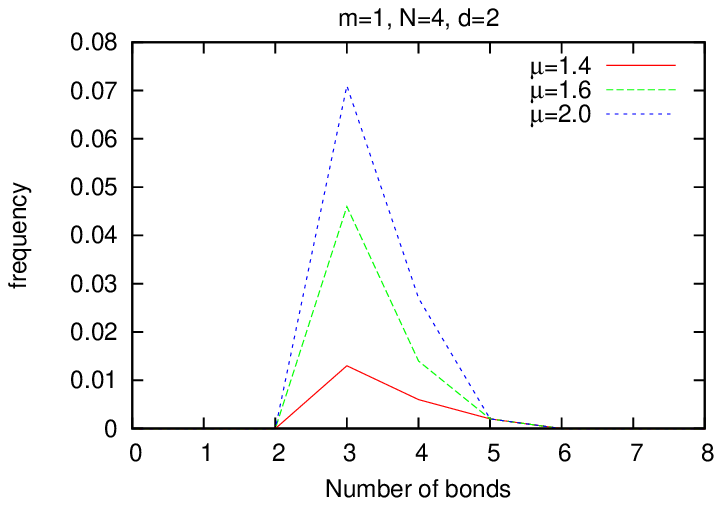}
\includegraphics[width=8cm,height=6cm]{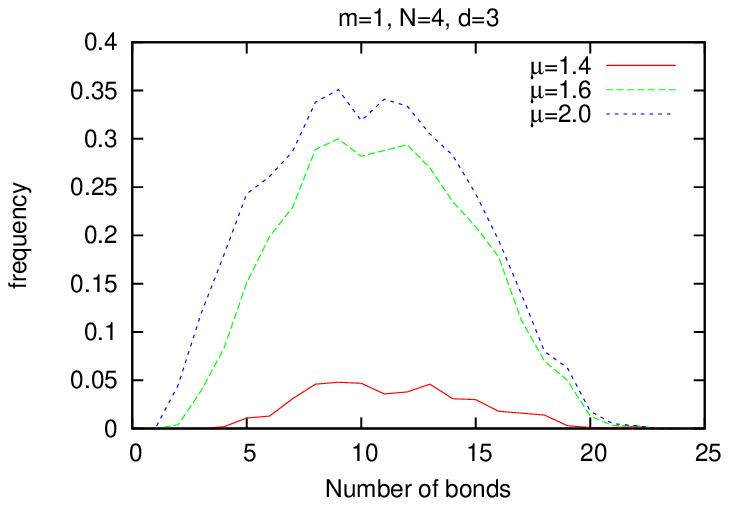}
\caption{Frequency of negative  $C$ where the configuration $k=(k_{x,\alpha})$ are chosen randomly.}\label{fig-1}
\end{figure}

If $d=1$, we can prove that for any $\mu>0$ and $m>0$, $
C(x_1,\cdots,x_{n}; D(\mu,m))>0$ for any number of different sites
$\{x_i\}_{i=1}^n$ (APPENDIX \ref{Appendix_2}). If $\mu=0$ and $m=0$, the fermion matrix $D$ is real and anti-Hermitian and thus its eigenvalues
comes in complex conjugate pairs with vanishing real part. If $\mu=0$ and $m>0$, the determinant of $D$ is positive. $D(\backslash \{x_1,\cdots,x_{2j}\})$ ($\mu=0$ and $m=0$) is also real and anti-Hermitian since the rows and columns corresponding to these sites are deleted. Thus if $\mu=0$ and $m>0$,
$C(x_1,\cdots,x_{n}; D(\mu,m))>0$ for any configuration $k$ and any dimension $d$. The numerical test
shows that the function $C$ in (\ref{2016_5_7_8}) for $d>1$, is
always positive for any configuration $k$ if $\mu\geq 0$ is close to
zero.  When $\mu$ is increased, $C$ may be negative for some
configurations. The left figure of FIG. \ref{fig-1} shows that the
frequency of negative $C$ for two dimensional Thirring model is
rather small, which is less than 0.1. For the three dimensional
Thirring model, the frequency of negative $C$ becomes larger (close
to 0.35 when $\mu=2$). Moreover, when $\mu$ is increased, the
frequency of negative $C$ also becomes larger. In fact, our
simulation shows that this frequency is zero when $\mu\leq 1.3$ for
both two and three dimensional Thirring model. Thus the presentation
of the partition function (\ref{2016_5_7_7}) avoid the sign problem
at least for small chemical potential.

The chiral condensate is
\begin{eqnarray}\label{2016_5_9_1}
\langle \bar \psi \psi \rangle  = \frac{1}{N^d}\frac{\partial \ln
Z}{\partial m} = \frac{1}{N^d}\Big\langle \frac{ \partial_m
C(x_1,\cdots,x_{2j})}{ C(x_1,\cdots,x_{2j})}\Big  \rangle
\end{eqnarray}
where the average is taken with respect to the weight of the
partition function (\ref{2016_5_7_7}). Similar to the calculation of
$C$ in (\ref{2016_5_7_8}), the ratio $\partial_m C/C$
have two formulae
\begin{eqnarray}\label{2016_5_9_2}
\frac{\partial_m C(x_1,\cdots,x_{2j})}{C(x_1,\cdots,x_{2j})} \nonumber &=& \sum_{x\neq x_1,\cdots,x_{2j}} \frac{\det
G(\{x, x_1,\cdots,x_{2j}\})}{\det G(\{x_1,\cdots,x_{2j}\})} \nonumber \\ &=&   \sum_{x\neq
x_1,\cdots,x_{2j}} \frac{\det D(\backslash \{x, x_1,\cdots,x_{2j}\})}{\det D(\backslash \{x_1,\cdots,x_{2j}\})}
\end{eqnarray}
The ratio between the determinant of submatrix $G$ can be obtained by
\begin{eqnarray}\label{2016_10_30_3}
   \frac{\det
G(\{x, x_1,\cdots,x_{2j}\})}{\det G(\{x_1,\cdots,x_{2j}\})}=
  G(\{x\}) - G(x,\textrm{occu\_sites})G(\{x_1,\cdots,x_{2j}\})^{-1}
G(\textrm{occu\_sites},x)
\end{eqnarray}
where $\textrm{occu\_sites}=(x_1,\cdots,x_{2j})$ denotes $2j$
occupied sites, $G(x,\textrm{occu\_sites})$ is a row vector with
$2j$ components, $D^{-1}_{x,x_i}$, $i=1,\cdots,2j$.
The column vector $G(\textrm{occu\_sites},x)$ is the transpose of $G(x,\textrm{occu\_sites})$.
The ratio between the determinant of submatrix $D$ is
\begin{eqnarray}\label{2016_10_30_5}
    \frac{\det D(\backslash \{x, x_1,\cdots,x_{2j}\})}{\det D(\backslash \{x_1,\cdots,x_{2j}\})}=
  \textrm{D\_Inv}(x,x)
\end{eqnarray}
where $  \textrm{D\_Inv}(x,x)$ is the diagonal element of $D(\backslash \{x_1,\cdots,x_{2j}\})^{-1}$ corresponding
to site $x\neq x_1,\cdots,x_{2j}$.

Similarly, the fermion density is
\begin{eqnarray}\label{2016_5_9_3}
\langle n  \rangle  = \frac{1}{N^d}\frac{\partial \ln Z}{\partial
\mu} = \frac{1}{N^d}\Big\langle \frac{ \partial_\mu
C(x_1,\cdots,x_{2j})}{ C(x_1,\cdots,x_{2j})}\Big  \rangle
\end{eqnarray}
can also be calculated.

The Monte Carlo algorithm based on the partition function in
(\ref{2016_5_7_7}) can be found in
Ref.\cite{Chandrasekharan_2011_5276}. We adopt the following three
steps to update the current configuration. Assume that the current
configuration $k$ has $n_b$ bonds
$$ C=([x_1,x_2],\cdots, [x_{2n_b-1},x_{2n_b}]) $$
Try to delete a bond, e.g. $[x_{2n_b-1},x_{2n_b}]$  from the current
configuration $C$ to be $$ C^\prime=([x_1,x_2],\cdots,
[x_{2n_b-3},x_{2n_b-2}])
$$
According to the detailed balance
\begin{eqnarray}\label{2016_5_9_8}
  W(C) P_{try}(C\rightarrow C^\prime) P_{acc}(C\rightarrow C^\prime) =
W(C^\prime) P_{try}(C^\prime\rightarrow C)
P_{acc}(C^\prime\rightarrow C)
\end{eqnarray}
where $W(C)$ and $W(C^\prime)$ are the weight in the partition
function (\ref{2016_5_7_7}) for the configuration $C$ and
$C^\prime$, respectively. The try probability from $C (C^\prime)$ to
$C^\prime (C)$ are
\begin{eqnarray*} P_{try}(C\rightarrow C^\prime) = \frac{1}{n_b}, \quad
P_{try}(C^\prime\rightarrow C) = \frac{1}{n_f}
\end{eqnarray*}
respectively.  Here $n_f$ is the number of bonds which can be
created from the configuration $C^\prime$. Thus accept probability
from $C$ to $C^\prime$ is
\begin{eqnarray*}
 P_{acc}(C\rightarrow C^\prime) = \frac{n_b}{n_f} \frac{W(C^\prime)}{W(C)}
\end{eqnarray*}
Try to add a bond, e.g. $ [x_{2n_b+1},x_{2n_b+2}] $ from the current
configuration $C$ to be
\begin{eqnarray*}
 C^\prime=([x_1,x_2],\cdots, [x_{2n_b-1},x_{2n_b}],[x_{2n_b+1},x_{2n_b+2}])
\end{eqnarray*}
The detailed balance is Eq. (\ref{2016_5_9_8}) where
\begin{eqnarray*}
P_{try}(C\rightarrow C^\prime) = \frac{1}{n_f}, \quad
P_{try}(C^\prime\rightarrow C) = \frac{1}{n_b+1}
\end{eqnarray*}
 Here $n_f$ is the
number of bonds which can be created from the configuration $C$.
Thus the accept probability from $C$ to $C^\prime$ is
\begin{eqnarray*}
 P_{acc}(C\rightarrow C^\prime) = \frac{n_f}{n_b+1} \frac{W(C^\prime)}{W(C)}
\end{eqnarray*}

Try to delete a bond, e.g. $ [x_{2n_b-1},x_{2n_b}] $  from the
current configuration $C$ and then add a bond, e.g., $
[y_{2n_b-1},y_{2n_b}] $
$$ C^\prime=([x_1,x_2],\cdots, [x_{2n_b-3},x_{2n_b-2}],[y_{2n_b-1},y_{2n_b}])
$$
In the detailed balance (\ref{2016_5_9_8}),
\begin{eqnarray*}  P_{try}(C\rightarrow C^\prime) = P_{try}(C^\prime\rightarrow C) =
\frac{1}{n_bn_f}
\end{eqnarray*} Here $n_f$ is the number of bonds which can be
created from the configuration $C$ where $ [x_{2n_b-1},x_{2n_b}] $
is deleted. Thus the accept probability to move a bond is
\begin{eqnarray*}
 P_{acc}(C\rightarrow C^\prime) =  \frac{W(C^\prime)}{W(C)}
\end{eqnarray*}

\section{Complex Langevin dynamics}\label{Langevin dynamics}
The expansion of (\ref{2016_5_7_6}) can also be written as an
integral of bosonic variables $A_{\alpha}(x)$ by
Hubbard-Stratonovich transformation
\begin{eqnarray}\label{2016_5_9_10}
&&  \exp\Big(U\sum_{x,\alpha=0,\cdots,d-1}
\bar\psi(x)\psi(x)\bar\psi(x+\hat\alpha)\psi(x+\hat\alpha)\Big) \nonumber \\
&=&\prod_{x,\alpha}\Big(\frac{1}{2\pi U}\Big)^{1/2}\int
\prod_{x,\alpha}dA_{\alpha}(x)
\exp\Big(-\frac{1}{8U}\sum_{x,\alpha}A_{\alpha}^2(x)\Big)  \\
& &  \exp\Big(-\sum_{x,y} \bar\psi(x) \sum_{\alpha}
i\frac{1}{2}\Big(B_{x,\alpha}A_{\alpha}(x) \delta_{x+\hat\alpha,y}+
C_{x-\hat\alpha,\alpha}A_{\alpha}(y)\delta_{x,y+\hat\alpha}\Big)
\psi(y) \Big)  \nonumber
\end{eqnarray}
for any two bosonic fields $B_{x,\alpha}$ and $C_{x,\alpha}$
satisfying $B_{x,\alpha}C_{x,\alpha}=1$.

Choosing
\begin{eqnarray}\label{2016_5_10_0}
  B_{x,\alpha} =e^{\mu\delta_{\alpha,0}} \eta_{x,\alpha}, \quad C_{x,\alpha}=
e^{-\mu\delta_{\alpha,0}} \eta_{x,\alpha}
\end{eqnarray}
and inserting (\ref{2016_5_9_10}) to the partition function $Z$ in
(\ref{2016_5_7_0}) and integrating the Grassmann fields $\psi$,
$\bar\psi$, one has
\begin{eqnarray}\label{2016_5_9_12}
 Z =  \int \prod_{x,\alpha}dA_{\alpha}(x)
\exp\Big(-\frac{1}{8U}\sum_{x,\alpha} A_{\alpha}^2(x)   \Big)\det K
=  \int \prod_{x,\alpha}dA_{\alpha}(x) e^{-S_{\textrm{eff}}}
\end{eqnarray}
where we omitted the factor $\prod_{x,\alpha}\Big(\frac{1}{2\pi
U}\Big)^{1/2}$.
 The matrix $K$ depends on
$A$
\begin{eqnarray}\label{2016_5_9_13}
K_{x,y} =  \sum_{\alpha=0,\cdots,d-1} \frac{\eta_{x,\alpha}}{2}\Big(
(s^{+}_{x,\alpha}+iA_\alpha(x)) e^{\mu \delta_{\alpha, 0}}
\delta_{x+\hat\alpha,y}-(s^{-}_{x,\alpha}-iA_\alpha(y))e^{-\mu
\delta_{\alpha, 0}}  \delta_{x,y+\hat\alpha}\Big)+ m\delta_{x,y}
\end{eqnarray}
which is complex, although the fields $A$ is real. The matrix $K$ is
reduced to be $D$ in (\ref{2016_5_7_2}) if $A_\alpha$ vanishes. The
effective action in (\ref{2016_5_9_12}) is
\begin{eqnarray}\label{2016_7_2_3} S_{\textrm{eff}} = \frac{1}{8U}\sum_{x,\alpha}
A_{\alpha}^2(x) -    \ln \det K
\end{eqnarray} The complex Langevin dynamics reads
\begin{eqnarray}\label{2016_5_9_15}
  A_{\alpha}(x,\Theta+\Delta \Theta) = A_{\alpha}(x,\Theta)  -\Delta t \frac{\partial S_{\textrm{eff}}}{\partial A_\alpha(x,\Theta)}
+ \sqrt{2\Delta t} \eta_\alpha(x,\Theta)
\end{eqnarray}
where $\Theta$ denotes the discrete complex Langevin time, $\Delta
\Theta$ is the time step. The real white noise $\eta_{x,\Theta}$
satisfies
$$
\langle \eta_{\alpha}(x,\Theta)
\eta_{\alpha^\prime}(x^\prime,\Theta^\prime) \rangle
=\delta_{\alpha,\alpha^\prime}
\delta_{x,x^\prime}\delta_{\Theta,\Theta^\prime}
$$
The drift force can be written as
\begin{eqnarray}\label{2016_5_9_16}
&& -\frac{\partial S_{\textrm{eff}}}{\partial A_{\alpha}(x)}
= -\frac{1}{4U} A_\alpha(x)
+\text{Tr} \Big(  K^{-1} \frac{\partial  K }{\partial A_\alpha(x)}
 \Big)\nonumber \\ &=&   -\frac{1}{4U} A_\alpha(x) + \frac{i}{2} \Big(
\eta_{x,\alpha} e^{\mu \delta_{\alpha,0}} K^{-1}_{x+\hat \alpha,x} +
\eta_{x+\hat\alpha,\alpha} e^{-\mu \delta_{\alpha,0} }
K^{-1}_{x,x+\hat \alpha} \Big)
\end{eqnarray}
The chiral condensate in (\ref{2016_5_9_1}) is written as
\begin{eqnarray} \label{2016_5_9_17}
 \langle \bar\psi\psi \rangle
= \frac{1}{N^d}\langle \textrm{Tr}(K^{-1} ) \rangle
\end{eqnarray}
and the fermion density in (\ref{2016_5_9_3}) reads
\begin{eqnarray}\label{2016_5_9_18}
 \langle n \rangle =
\frac{1}{N^d}\Big\langle \textrm{Tr}(K^{-1} \frac{\partial
K}{\partial \mu} ) \Big\rangle
\end{eqnarray}
where the average is taken with respect to weight
$e^{-S_{\textrm{eff}}}$. Note that
\begin{eqnarray*}
\textrm{Tr}(K^{-1} \frac{\partial K}{\partial \mu} )     = \sum_{x,y
}K^{-1}_{y,x}  \Big( \frac{e^\mu }{2}
 (s^{+}_{x,0}+iA_0(x))\delta_{x+\hat 0,y}+\frac{e^{-\mu} }{2}(s^{-}_{x,0}-iA_0(y))\delta_{x,y+\hat 0} \Big)
\end{eqnarray*}

If we can choose instead of (\ref{2016_5_10_0})
\begin{eqnarray*}
 B_{x,\alpha} =e^{\mu\delta_{\alpha,0}} \eta_{x,\alpha}s^{+}_{x,\alpha}, \quad C_{x,\alpha}=
e^{-\mu\delta_{\alpha,0}} \eta_{x,\alpha}s^{-}_{x+\hat\alpha,\alpha}
\end{eqnarray*}
satisfying $B_{x,\alpha}C_{x,\alpha}=1$, the partition function $Z$
can also be written as Eq. (\ref{2016_5_9_12}), where the matrix $K$
is replaced by
\begin{eqnarray}\label{2016_5_9_13_1}
\tilde K_{x,y} &=&  \sum_{\alpha=0,\cdots,d-1}
\frac{\eta_{x,\alpha}}{2}\Big( s^{+}_{x,\alpha}(1+iA_\alpha(x)) e^{\mu
\delta_{\alpha, 0}} \delta_{x+\hat\alpha,y} \nonumber
\\ && -s^{-}_{x,\alpha}(1-iA_\alpha(y))e^{-\mu \delta_{\alpha, 0}}
\delta_{x,y+\hat\alpha}\Big)+ m\delta_{x,y}
\end{eqnarray}

\section{Simulation results}\label{results}
The implementation of fermion bag approach and complex Langevin
dynamics can be found in \cite{Daming_0710.0323}. We use the
$\Gamma$ method to estimate the error for the samples in each Monte
Carlo simulation or complex Langevin dynamics \cite{Wolff.143}. The
following simulation results are given for one and three dimensional
Thirring model with fixed $N=8$, $m=1$ but with different coupling
strength $U$ and different chemical potential $0\leq\mu\leq 2$.

\begin{figure}
\centering
\includegraphics[width=8cm,height=6cm]{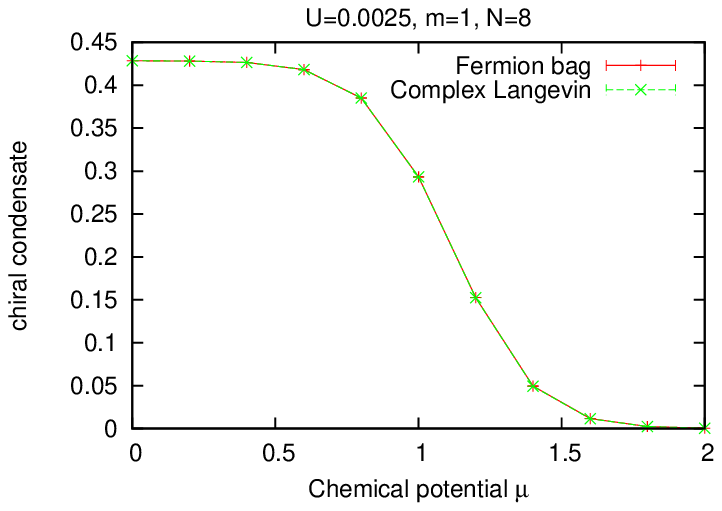}
\includegraphics[width=8cm,height=6cm]{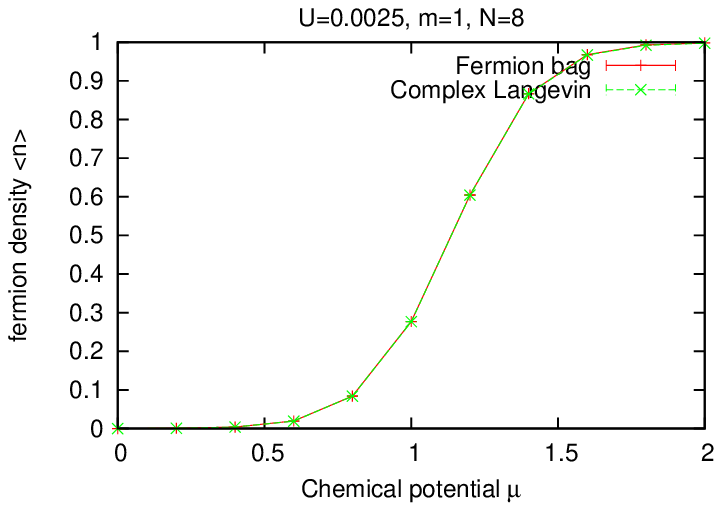}
\caption{Comparison of chiral condensate and fermion density
obtained by complex Langevin dynamics and by fermion bag approach.
In fermion bag approach, the sampling starts after $1\times 10^6$
Monte Carlo step and finished after $1\times 10^7$ steps. Two
subsequent samples are separated by $10\times 3\times N^3$ Monte
Carlo steps. In complex Langevin dynamics, $\Delta\Theta=0.001$, the
equilibrium Langevin time $t_{\textrm{eq}}=20$, the sampling end
time $t_{\textrm{end}}=40$. Two subsequent samples are separated by
10 complex Langevin steps.}\label{fig0}
\end{figure}

\begin{figure}
\centering
\includegraphics[width=8cm,height=6cm]{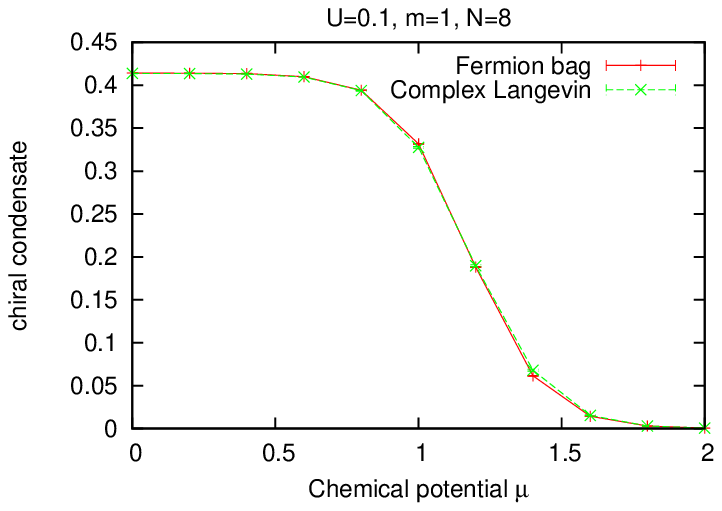}
\includegraphics[width=8cm,height=6cm]{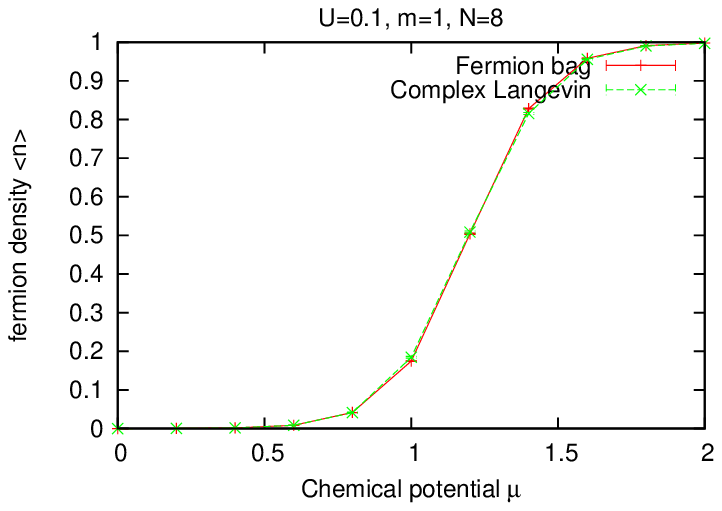}
\caption{Parameters are the same with those in FIG. \ref{fig0},
except $U$.}\label{fig1}
\end{figure}

\begin{figure}
\centering
\includegraphics[width=8cm,height=6cm]{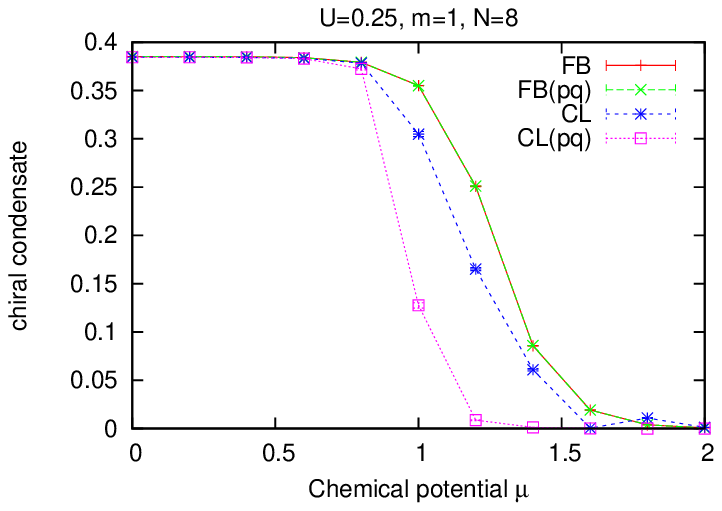}
\includegraphics[width=8cm,height=6cm]{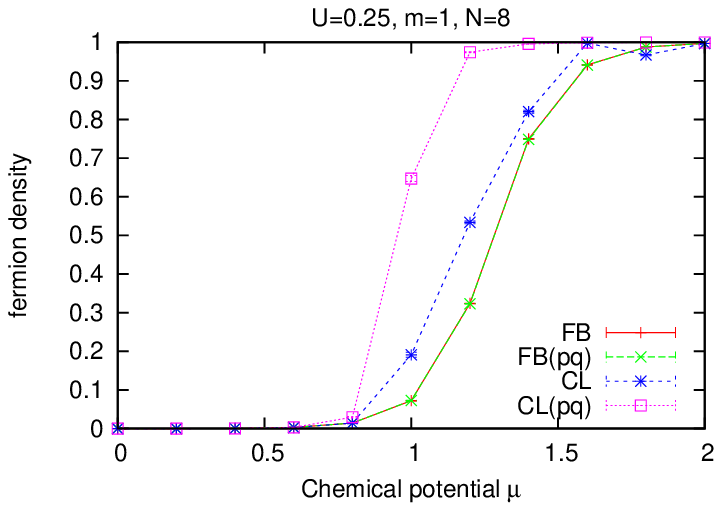}
\caption{Parameters are the same with those in FIG. \ref{fig0},
except $U$. FB-Fermion Bag, CL-Complex Langevin, pq-phase
quenched.}\label{fig3}
\end{figure}

\begin{figure}
\centering
\includegraphics[width=8cm,height=6cm]{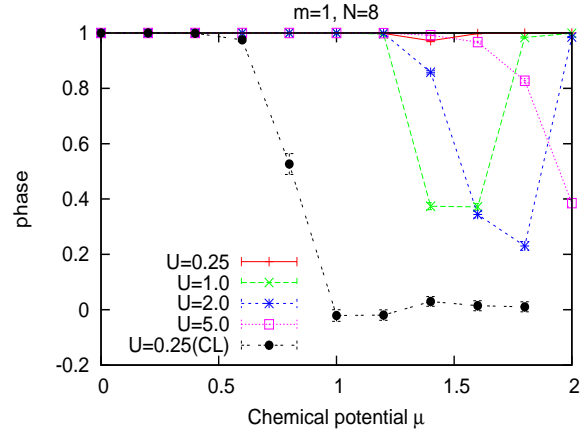}
\caption{The real part of phase $\langle e^{i\varphi}\rangle_{\text{pq}}$
obtained by fermion bag approach for $U=0.25,1.0,2.0,5.0$ and by
complex Langevin dynamics for $U=0.25$ (U=0.25(CL)).}\label{fig6}
\end{figure}

FIG.\ref{fig0} and FIG.\ref{fig1} shows the comparison of the chiral
condensate and fermion density obtained by fermion bag approach (FB) and
by complex Langevin dynamics (CL) for different chemical potential $\mu$
and coupling strength $U$. Both these averages agree with each other
very well by these two numerical methods. The statistic error are almost invisible in FIG.\ref{fig0} and FIG.\ref{fig1}.
 When the coupling strength
$U$ is increasing, e.g., $U=0.25$, the chiral condensate and fermion
density obtained by FB and by CL are quite different for the intermediate
values of chemical potential $\mu$, as shown in
FIG.\ref{fig3}. One reason of this difference is related to the severeness of the sign problem, which can be measured by the phase $\langle e^{i\varphi}\rangle_{\text{pq}} =
Z/Z_{\text{pq}}$. The sign problem is rather severe
for CL, while it is still mild for FB if $1\leq \mu \leq 2$. We thus compare the results obtained by these two
numerical methods under the phase quenched approximation (APPENDIX
\ref{Appendix_3}). The chiral condensate and fermion density agree
with each other for FB and this method under
phase quenched approximation (FB(pq)). While these agreement can not
be achieved for CL and the complex Langevin
dynamics under phase quenched approximation (CL(pq)). The severity
of the sign problem by both approaches is shown in FIG.\ref{fig6}.
Because the determinant $\det (K)$ of $K$ becomes too large if
$\mu>1.6$, we just calculate the phase $\langle
e^{i\varphi}\rangle_{\text{pq}}$
 by CL for $0\leq\mu\leq 1.6$. We also
 calculate this phase by FB for different $U$ and
 different
 chemical potential $0\leq \mu\leq 2$. For $U=0.25$, the phase $\langle
 e^{i\varphi}\rangle_{\text{pq}}$ is almost very close to 1 for FB
in $0\leq\mu\leq 2$. Thus the sign problem is almost overcome and
this can explain why the results obtained
 by FB agrees with those obtained under the quenched approximation (FB(pq)) (See FIG.\ref{fig3}).
     The sign problem for FB becomes severe
 during the intermediate range of chemical potential, as shown
 for $U=0.25,1.0,2.0$. Moreover this range shift to larger
 chemical potential if $U$ is increased. For CL  with $U=0.25$, the
 phase drops very fast to zero if $\mu>0.6$ and is very close to
 zero for $\mu\geq 1$. This also explain the difference between
 those obtained by CL and by CL(pq) in FIG.\ref{fig3}.

\begin{figure}
\centering
\includegraphics[width=8cm,height=6cm]{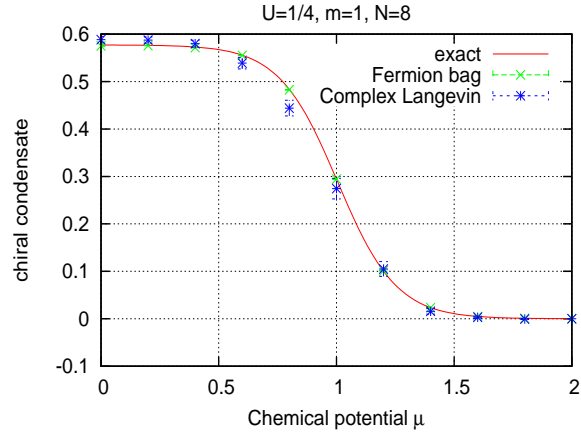}
\caption{Comparison between Fermion bag approach, complex Langevin
dynamics and exact solution for one dimensional Thirring model (See Ref.\cite{Daming_0710.0323}).}\label{fig7}
\end{figure}

As shown in FIG.\ref{fig6}, the (real part of) phase drops rapidly in the intermediate value of $\mu$ ($0.6\leq\mu\leq 1.2$) for CL (U=0.25(CL)).
Although the statistical error of the chiral condensate and fermion density in this range of $\mu$
is larger than those for $\mu<0.6$ or $\mu>1.2$, the statistical error in the whole value of $\mu$ is
almost invisible in FIG.\ref{fig3}. In
FIG.\ref{fig7}, we compared the chiral condensate obtained by FB and by CL with the exact result for one dimensional Thirring model with the same parameters \cite{Daming_0710.0323}. The chiral condensate obtained by FB agrees with the exact result in the whole range of $\mu$,
while the chiral condensate by CL is slightly smaller than the exact result in the intermediate value of $\mu$, where the phase $Z/Z_{\text{pq}}$ drops rapidly from 1 to 0 for CL. These results is quite similar with those in the left figure of FIG.\ref{fig3}, where the chiral condensate obtained by CL is smaller than those obtained by FB in the intermediate value of $\mu$. The statistical error for CL in left
figure of FIG.\ref{fig3} is smaller than those in FIG.\ref{fig7}.

Our calculation for the chiral condensate and fermion density in
one and three dimensional Thirring model at finite density by CL quantitatively agree with those obtained by Pawlowski etc. \cite{Pawlowski_2013_2249}\cite{Pawlowski_2013_094503}. Compared with the lower figure of
  Figure 3 in Ref.\cite{Pawlowski_2013_2249}, where $\beta=1=1/(4U)$, i.e., $U=0.25$, our result by CL is more close to those obtained by FB
  in the left figure of FIG.\ref{fig3}. When $\mu=1$, the chiral condensate are $0.305\pm 0.0027$ by CL and $0.356\pm 0.00011$ by FB, respectively in the left figure of FIG.\ref{fig3} while it is 0.25 in lower figure of
  Figure 3 of Ref.\cite{Pawlowski_2013_2249}. The statistical error is also almost invisible in
   Figure 3 of Ref.\cite{Pawlowski_2013_2249} in the intermediate value of $\mu$ where the phase drops rapidly in this range as shown in
 Figure 4 of Ref.\cite{Pawlowski_2013_2249}. We can also compare the chiral condensate of one dimensional Thirring model in FIG.\ref{fig7} with
 Figure 5(b) in Ref.\cite{Pawlowski_2013_094503}. Our result in FIG.\ref{fig7} by CL is better than those in Figure 5(b) of \cite{Pawlowski_2013_094503}.
 For example, at $\mu=1$, the chiral condensate obtained by CL is $0.27\pm 0.03$ and the exact is $0.293$ in FIG.\ref{fig7}, while it is
 $0.14\pm0.023$ in Figure 5(b) in Ref.\cite{Pawlowski_2013_094503}. Moreover the statistical error in Figure 5(b) of Ref.\cite{Pawlowski_2013_094503}
 are larger than those (e.g., Figure 3 in Ref.\cite{Pawlowski_2013_094503}) in three dimensional Thirring model at finite density,
 which is quite similar to the statistical error in our calculation by CL for one and three dimensional Thirring model.

The discussion above shows that the difference of chiral condensate obtained by CL and by FB in the intermediate value of $\mu$
is definitely related to the fast decay of real part of phase $\langle e^{i\varphi}\rangle_{\text{pq}}$, i.e., the severity of the sign problem,
although the statistical error is small as shown in FIG.\ref{fig3}. According to Ref. \cite{Pawlowski_2013_094503}\cite{Aarts_2011_3270}, the quantity
\begin{eqnarray}\label{2016_10_30}
 \langle LO\rangle\equiv \Big\langle \sum_{x,\mu} \Big(\frac{d}{dA_\mu(x)} - \frac{dS_{\text{eff}}}{dA_\mu(x)}\Big)\frac{d}{dA_\mu(x)} O(A)
\Big\rangle
\end{eqnarray}
should vanish for any holomorphic function $O(A)$ if CL works. We choose the observable (the chiral condensate) $O(A)=\frac{1}{N^d}\text{Tr}(K^{-1})$ for $\mu=1$, $m=1$ and $N=8$. In the one dimensional case, $\langle LO\rangle$ is $0.0137\pm 0.00708$ if $U=0.0025$ and becomes $-5.88\pm 7.33$ if $U=0.16$. In the three dimensional case, $\langle LO\rangle$ is $-1.207\pm 0.0025$ if $U=0.0025$, $61.6\pm 64.02$ if $U=0.1$ and $-206.3\pm 420.7$ if $U=0.25$. Thus $\langle LO\rangle$ becomes large if $U$ is increased and the chiral condensate by CL for $\mu=1$ in the left figure of FIG.\ref{fig3} is not reliable.

  Finally we also compared the chiral condensate obtained by FB and by CL for
one dimensional Thirring model with parameters $U=10$, $m=1$ and $N=8$ (Figure 5 in \cite{Daming_0710.0323}). FB recover the exact result for
large coupling strength $U=10$ for the chemical potential $0\leq\mu\leq 2$ while the result obtained by CL is totally wrong.
This is because there is no sign problem in FB in one dimensional Thirring model, while the sign problem is very severe in CL.

\begin{figure}
\centering
\includegraphics[width=8cm,height=6cm]{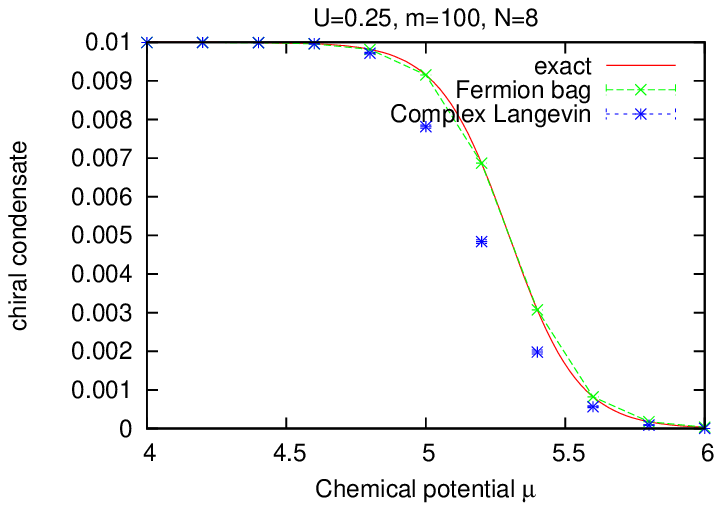}
\includegraphics[width=8cm,height=6cm]{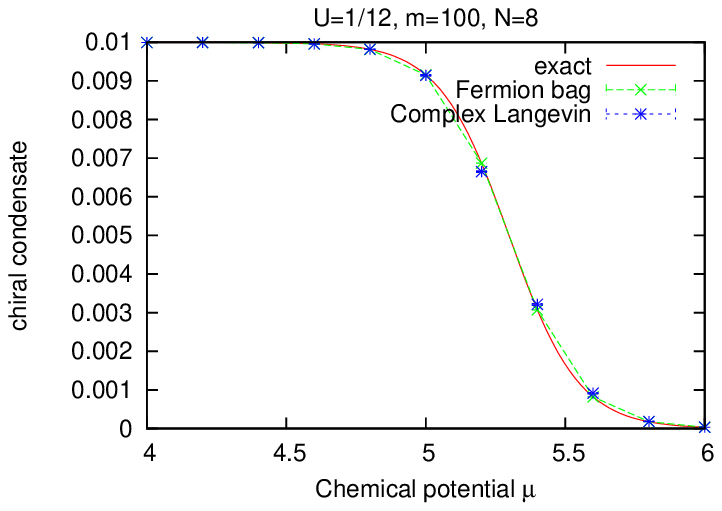}
\caption{The chiral condensate in the heavy quark limit for
 coupling strength $U=0.25$ (Left) and $U=1/12=0.08333$ (Right). Parameters are the same with those
in FIG. \ref{fig0}, except $U$ and $m$.}\label{fig4}
\end{figure}

In the heavy fermion limit
$$ m\rightarrow \infty , \quad \mu \rightarrow \infty , \quad \zeta \equiv  (2m)^{-1} e^\mu  \ \textrm{fixed} $$
the exact solution is known \cite{Pawlowski_2013_2249}, which does
not depend on $U$ (APPENDIX \ref{Appendix_1.1}). FIG.\ref{fig4}
shows the comparison between the condensate calculated by FB and by CL with the exact
solution for different coupling strength $U$ in this limit. The
results obtained by FB agree with the exact result
for the different coupling strength $U$. The results obtained by
CL agree with the exact result only when $U$
is small, e.g. $U=1/12$. When $U$ is increased, e.g., $U=0.25$, the
chiral condensate obtained by CL is less than
the exact result in the intermediate range of chemical potential
$4.8\leq\mu\leq 5.6$, which was also found in
Ref.\cite{Pawlowski_2013_2249}.

\section{Conclusions}\label{conclusion}
The three dimensional massive Thirring model at finite density are
solved by two numerical methods: fermion bag approach and complex
Langevin dynamics. Two average quantities, chiral condensate and
fermion density, are calculated and are compared by these numerical
methods. If the fermion coupling strength $U$ is small, these
averages obtained by fermion bag approach agree with those obtained
by complex Langevin dynamics. When $U$ and chemical potential are
increasing, the sign problem for complex Langevin becomes severe,
the results obtained by complex Langevin dynamics are quite
different with those obtained under the phase quenched
approximation. For the parameters, where the sign problem becomes
severe for complex Langevin dynamics, the sign problem for the fermion bag approach is
still mild and thus the result obtained by fermion
bag approach are reliable for these model parameters. Moreover, in
the heavy quark limit, the fermion bag approach can recover the
exact result for large coupling strength $U$, while the complex
Langevin dynamics just recover the exact result for small coupling
strength $U$. I believe that these advantages of the fermion bag
approach over complex Langevin dynamics can be checked for the other
interacting fermion systems with finite density, e.g., Gross-Neveu
model, Yukawa model, etc.

\vspace{1cm}

 Acknowledgments.
  I would like to thank Prof. Shailesh
Chandrasekharan for discussion. Daming Li was supported by the
National Science Foundation of China (No. 11271258, 11571234).

{}

\appendix

\section{\label{Appendix_1}Proof of (\ref{2016_5_8_0})}
For any different sites $\{x_i\}_{i=1}^n$,
\begin{eqnarray}\label{2016_5_8_30}
&& C(x_1,\cdots,x_{n}; D(\mu,m))  \nonumber \\ &=& \int d\bar\psi
d\psi \exp\Big(-\sum_{x,y} \bar\psi(x)D(\mu,m)_{x,y}\psi(y)\Big)
\bar\psi(x_1)\psi(x_1)\cdots \bar\psi(x_{n})\psi(x_{n}) \nonumber \\
&=&  \int d\bar\chi d\chi \exp\Big(-\sum_{x,y}
\chi(x)D(\mu,m)_{x,y}\bar\chi(y)\Big)  \chi(x_1)\bar\chi(x_1)\cdots
 \chi(x_{n})\bar\chi(x_{n}), \quad \bar\psi \rightarrow  \chi, \ \psi\rightarrow \bar\chi \nonumber \\
&=& \int d\bar\chi d\chi \exp\Big(\sum_{x,y}
 \chi(x)D(-\mu,-m)_{y,x}\bar\chi(y)\Big) \chi(x_1)\bar\chi(x_1)\cdots
 \chi(x_{n})\bar\chi(x_{n})\quad \textrm{ by } (\ref{2016_5_7_4})\nonumber \\
 &=& (-1)^n\int d\bar\chi d\chi \exp\Big(-\sum_{x,y}
 \bar\chi(x)D(-\mu,-m)_{x,y}\chi(y)\Big) \bar\chi(x_1)\chi(x_1)\cdots
\bar\chi(x_{n})\chi(x_{n})\nonumber \\
& = & (-1)^n C(x_1,\cdots,x_{n}; D(-\mu,-m))
\end{eqnarray}
In the second equality we used $d\bar\psi d\psi= d\chi
d\bar\chi=d\bar\chi d\chi$ since there are even number of sites. By
the symmetry (\ref{2016_5_7_5}) of $D$,
\begin{eqnarray}\label{2016_5_8_31}
&& C(x_1,\cdots,x_{n}; D(\mu,m)) \nonumber \\ &=& \int d\bar\psi
d\psi \exp\Big(\sum_{x,y}
\bar\psi(x)\varepsilon_xD(\mu,-m)_{x,y}\varepsilon_y\psi(y) \Big)
\bar\psi(x_1)\psi(x_1)\cdots \bar\psi(x_{n})\psi(x_{n}) \nonumber \\
&=& (-1)^n\int d\bar\chi d\chi \exp\Big(- \sum_{x,y}\bar\chi(x)
D(\mu,-m)_{x,y}\chi(y)\Big) \bar\chi(x_1)\chi(x_1)\cdots
\bar\chi(x_{n})\chi(x_{n})   \nonumber  \\
&=& (-1)^n C(x_1,\cdots,x_{n}; D(\mu,-m))
\end{eqnarray}
where in the second equality we used $
\bar\psi(x)=-\varepsilon_x\bar\chi(x)$, $\psi(x) = \varepsilon_x
\chi(x)$, and thus $ d\bar\psi d\psi= d\bar\chi d\chi$ due to even
number of sites. Combing (\ref{2016_5_8_30}) and
(\ref{2016_5_8_31}), we obtain (\ref{2016_5_8_0}).

\section{\label{Appendix_2}There is no sign problem for $d=1$}
If $d=1$, the $N\times N$ (even $N$) fermion matrix is
\begin{eqnarray*}
D = D(\mu,m) = \left( \begin{array}{cccccc}
m  &  \frac{e^\mu}{2} &  &  &  & \frac{e^{-\mu}}{2}  \\
-\frac{e^{-\mu}}{2} & m  &  \frac{e^{\mu}}{2}   &  &  & \\
& -\frac{e^{-\mu}}{2} & m  &  \frac{e^{\mu}}{2}    &  &  \\
&  &   &   \ddots   &  &  \\
    &  &   &  & m &  \frac{e^{\mu}}{2} \\
-\frac{e^{\mu}}{2}  &   & &   & -\frac{e^{-\mu}}{2}  & m   \\
\end{array} \right)_{N\times N}
\end{eqnarray*}
According to a formula of the determinant \cite{Molinari.2221}, the
determinant of $D$ is
\begin{eqnarray*}
\det D & =&      \frac{e^{N\mu}}{2^N}  + \frac{e^{-N\mu}}{2^N}  +
\textrm{Tr} (T)
\end{eqnarray*}
where the $2\times 2$ transfer matrix $T$ is $ T  =  \left(
\begin{array}{cc}
m &  \frac{1}{4}  \\
1 & 0
\end{array} \right)^N
$. Obviously, $\det D>0$ for any $\mu>0$ and $m>0$. Choose $n$
different indices, $1\leq i_1<\cdots<i_n\leq N$ and  delete $n$ rows
and columns corresponding to these $n$ indices from $D$ to obtain
$\tilde D$. We want to prove that $(N-n)\times (N-n)$ matrix $\tilde
D$ satisfies $\det \tilde D>0$. This holds because the structure of
$\tilde D$ is the same with $D$ and thus the determinant of $\tilde
D$ can be calculated \cite{Molinari.2221}, which must be positive.
For example, $N=10$, $n=2$, $i_1=4$, $i_2=7$,
\begin{eqnarray*}
\tilde D =   \left( \begin{array}{ccccccccccc}
*  &  * &  & | & & & | & &  & * \\
* &  *  & * & | &  & & | & &  \\
& * &  *  &  | &  &  & | &  & & \\
-& - &  -  & - & - & - & - & - & - & - \\
 &  &    & |  & *  & * & | &  &  &  \\
 &  &    & |  & *  & * & | &  &  &  \\
-& - &  -  & - & - & - & - & - & - & - \\
&  &    & | &  &  & | & * & * &  \\
&  &    & | &  &  & | & * & * & * \\
* &  &    & | &  &  & | &  & * & * \\
\end{array} \right)
\end{eqnarray*}
Since $C(x_1,\cdots,x_{2j})$ can be presented by the determinant of
the submatrix of $D$, which is nonnegative, the sign problem is
avoided for $d=1$.

\section{\label{Appendix_1.1} Heavy quark limit}
Introducing notations $X=(x_1,\cdots,x_{d-1})$,
$Y=(y_1,\cdots,y_{d-1})$. The matrix element of $\tilde K$ in
(\ref{2016_5_9_13_1}) can be written as
\begin{eqnarray*}
&&   \tilde K_{(t,X),(t,Y)} \equiv
  (B_t)_{X,Y} \\ & = & \sum_{\alpha=1,\cdots,d-1}
\frac{\eta_{x,\alpha}}{2}\Big(
 (1+iA_\alpha(x))
\delta_{x+\hat\alpha,y}- (1-iA_\alpha(y))
\delta_{x,y+\hat\alpha}\Big)+ m\delta_{x,y}, \quad t=0,\cdots,N-1
\end{eqnarray*}
\begin{eqnarray*}
  \tilde K_{(t,X),(t+1,Y)} = s^{+}_{x,0}  \frac{e^\mu}{2}
  (C_t)_{X,Y},  \quad   (C_t)_{X,Y} \equiv
   (1+iA_0(x))
\delta_{X, Y}   , \quad t=0,\cdots,N-1
\end{eqnarray*}
\begin{eqnarray*}
 \tilde K_{(t,X),(t-1,Y)} =
  -s^{-}_{x,0}\frac{e^{-\mu
 }}{2}    (C_{t-1}^*)_{X,Y}, \quad
t=0,\cdots,N-1
\end{eqnarray*}
The matrix $\tilde K$ is a $N^d\times N^d$ matrix
\begin{eqnarray*}
\tilde K = \left( \begin{array}{cccccc}
B_0  &  \frac{e^\mu}{2} C_0 &  &  &  & \frac{e^{-\mu}}{2} C_{N-1}^* \\
-\frac{e^{-\mu}}{2} C_{0}^* & B_1  &  \frac{e^{\mu}}{2} C_1  &  &  & \\
& -\frac{e^{-\mu}}{2} C_{1}^* & B_2  &  \frac{e^{\mu}}{2}  C_2  &  &  \\
&  &   &   \ddots   &  &  \\
    &  &   & -\frac{e^{-\mu}}{2} C_{N-3}^* & B_{N-2} &  \frac{e^{\mu}}{2} C_{N-2} \\
-\frac{e^{\mu}}{2} C_{N-1}  &   & &   & -\frac{e^{-\mu}}{2} C_{N-2}^* & B_{N-1}  \\
\end{array} \right)_{N^d\times N^d}
\end{eqnarray*}
 In the heavy quark limit
$$ m\rightarrow \infty , \quad \mu \rightarrow \infty , \quad \zeta \equiv  (2m)^{-1} e^\mu  \ \textrm{fixed} $$
The matrix $\tilde K$ becomes
\begin{eqnarray*}
(2m)^{-1}\tilde K_{x,y} &=&   \zeta
(s^{+}_{x,0}+iA_0(x))\delta_{x+\hat
0,y} + \delta_{x,y} \nonumber \\
&=&\left( \begin{array}{cccccc}
I  &  \zeta C_0 &  &  &  & 0 \\
0 & I  &  \zeta C_1  &  &  & \\
& 0 & I  &  \zeta  C_2  &  &  \\
&  &   &   \ddots   &  &  \\
    &  &   & 0 & I &  \zeta C_{N-2} \\
-\zeta C_{N-1}  &   & &   & 0 & I  \\
\end{array} \right)_{N^d\times N^d}
\end{eqnarray*}
The determinant of $\tilde K$ satisfies
$$ \frac{1}{(2m)^{N^d}} \det  \tilde K =  \det (I + \xi C_{N-1}C_0\cdots C_{N-2} ) =  \prod_X (1+\xi {\cal P}_X) $$
where $\xi = \zeta^N$ and ${\cal P}_X=\prod_t (1+A_0(t,X)) $ is the
Polyakov loop starting and ending at the space point $X$. The
partition function $Z$ in (\ref{2016_5_9_12}) reads
\begin{eqnarray*}
 Z &=&  (2m)^{N^d}  \int \prod_{x,\alpha}dA_{\alpha}(x)
\exp\Big(-\frac{1}{8U}\sum_{x,\alpha} A_{\alpha}^2(x) \Big)\prod_X
(1+\xi {\cal P}_X) \\
 &=&  (2m)^{N^d}  \Big(\frac{1}{2\pi U}\Big)^{\frac{(d-1)N^d}{2}} \int \prod_{x}dA_{0}(x)
\exp\Big(-\frac{1}{8U}\sum_{x} A_{0}^2(x) \Big)\prod_X
(1+\xi {\cal P}_X) \\
 &=&  (2m)^{N^d}  \Big(\frac{1}{2\pi U}\Big)^{\frac{(d-1)N^d}{2}} \prod_X \int \prod_{t}dA_{0}(t,X)
\exp\Big(-\frac{1}{8U}\sum_{t} A_{0}^2(t,X) \Big)
(1+\xi {\cal P}_X) \\
 &=&  (2m)^{N^d}  \Big(\frac{1}{2\pi U}\Big)^{\frac{dN^d}{2}}
 (1+\xi)^{N^{d-1}}
\end{eqnarray*}
where in the last equality we used
\begin{eqnarray*}
 && \int \prod_{t}dA_{0}(t,X)
\exp\Big(-\frac{1}{8U}\sum_{t} A_{0}^2(t,X) \Big) (1+\xi {\cal
P}_X) \\
  &=& \Big(\frac{1}{2\pi U}\Big)^{\frac{N}{2}} + \xi  \int \prod_{t}dA_{0}(t,X)
\exp\Big(-\frac{1}{8U}\sum_{t} A_{0}^2(t,X) \Big) {\cal P}_X  \\
  &=& \Big(\frac{1}{2\pi U}\Big)^{\frac{N}{2}} + \xi  \prod_t \int  dA_{0}(t,X)
\exp\Big(-\frac{1}{8U} A_{0}^2(t,X) \Big)  (1+A_0(t,X))\\
  &=& \Big(\frac{1}{2\pi U}\Big)^{\frac{N}{2}} + \xi  \Big(\frac{1}{2\pi U}\Big)^{\frac{N}{2}}
\end{eqnarray*}
In the heavy quantum limit, the chiral condensate and fermion
density are
 $$ \langle\bar\psi \psi \rangle  = \frac{1}{m(1+\xi)},
\quad  \langle n \rangle  = \frac{1}{1+\frac{1}{\xi} } $$
respectively, which does not depend on $U$
\cite{Pawlowski_2013_2249}.

\section{\label{Appendix_3}Phase quenched approximation}
The phase quenched approximation to (\ref{2016_5_9_12}) is to
replace $\det K$ by its module $|\det K| = \sqrt{\det(KK^\dagger)}$
\begin{eqnarray}\label{2016_6_29_1}
 Z_{\textrm{pq}} =  \int \prod_{x,\alpha}dA_{\alpha}(x)
\exp\Big(-\frac{1}{8U}\sum_{x,\alpha} A_{\alpha}^2(x)   \Big)|\det
K| =  \int \prod_{x,\alpha}dA_{\alpha}(x) e^{-S_{\textrm{pq,eff}}}
\end{eqnarray}
where the effective action is
\begin{eqnarray}\label{2016_7_2_1} S_{\textrm{pq,eff}} = \frac{1}{8U}\sum_{x,\alpha}
A_{\alpha}^2(x) -   \frac{1}{2} \ln \det (KK^\dagger)
\end{eqnarray}
Since
\begin{eqnarray*}
 && \frac{\partial}{\partial A_\alpha(x)}  \frac{1}{2} \ln \det
 (KK^\dagger)
  =  \frac{1}{2} \frac{\partial}{\partial A_\alpha(x)}   \text{Tr} \ln (KK^\dagger) \\
  &=& \frac{1}{2}    \text{Tr} \Big( (KK^\dagger)^{-1} \frac{\partial  (KK^\dagger)}{\partial A_\alpha(x)}   \Big)\\
  &=& \frac{1}{2}    \text{Tr} \Big( (KK^\dagger)^{-1} \Big[ \frac{\partial  K }{\partial A_\alpha(x)} K^\dagger
  + K\frac{\partial  K^\dagger }{\partial A_\alpha(x)}   \Big]  \Big)\\
    &=& \frac{1}{2}    \text{Tr} \Big(  K^{-1}  \frac{\partial  K }{\partial A_\alpha(x)}
  + K^{\dagger -1}\frac{\partial  K^\dagger }{\partial A_\alpha(x)}   \Big)\\
      &=&  \text{Re}  \Big[  \text{Tr} \Big(  K^{-1}  \frac{\partial  K }{\partial A_\alpha(x)}
 \Big)\Big]
\end{eqnarray*}
the drift force is
\begin{eqnarray*}
-\frac{\partial S_{\textrm{pq,eff}}}{\partial A_{\alpha}(x)}   =
-\frac{1}{4U} A_\alpha(x) +
 \text{Re}  \Big[  \text{Tr} \Big(  K^{-1}  \frac{\partial  K }{\partial A_\alpha(x)}
 \Big)\Big]
\end{eqnarray*}
which is just the real part of the complex drift force in
(\ref{2016_5_9_16}). This is because the effective action
(\ref{2016_7_2_1}) in the quenched approximation is taken to be the
real part of the complex effective action in (\ref{2016_7_2_3})
$$  e^{-S_{\text{eff}}} =  e^{-S_{\text{pq,eff}}} e^{i\varphi}, \quad e^{i\varphi} = \frac{\det K}{|\det K|} \quad \Longleftrightarrow \quad
   \text{Re}(\ln \det K) = \frac{1}{2} \ln \det (KK^\dagger) $$
Here the logarithm $\ln$ is understood to be the principal value of
the logarithm.

The chiral condensate in (\ref{2016_5_9_17}) and fermion density in
(\ref{2016_5_9_18}) is replaced by
$$
 \langle \bar\psi\psi \rangle = \frac{1}{N^d}\frac{\partial \ln
Z_{\textrm{pq}}}{\partial m} = \frac{1}{N^d}\langle \text{Re}
\textrm{Tr}(K^{-1} ) \rangle_{\text{pq}}, \quad
 \langle n \rangle = \frac{1}{N^d}\frac{\partial \ln
Z_{\textrm{pq}}}{\partial \mu}  = \frac{1}{N^d}\Big\langle \text{Re}
\textrm{Tr}(K^{-1} \frac{\partial K}{\partial \mu} )
\Big\rangle_{\text{pq}}
$$
where the average is taken with respect to the weight of partition
function in (\ref{2016_6_29_1}). The average phase factor in the
phase-quenched theory $\langle e^{i\varphi}\rangle_{\text{pq}} =
Z/Z_{\text{pq}}$ indicates the severeness of the sign problem in the
thermodynamic limit.

Since the real function $C$ may be negative, the phase quenched
approximation of (\ref{2016_5_7_7}) is
\begin{eqnarray}\label{2016_7_2_5}
 Z_{\textrm{pq}} = \sum_{k=(k_{x,\alpha})}   U^{j}|C(x_1,\cdots,x_{2j})|
\end{eqnarray}
The chiral condensate and fermion density under this quenched phase
approximation are
$$
 \langle \bar\psi\psi \rangle = \frac{1}{N^d}\frac{\partial \ln
Z_{\textrm{pq}}}{\partial m} = \frac{1}{N^d}\Big\langle
\frac{\partial_m C}{C}\Big \rangle_{\text{pq}}, \quad
 \langle n \rangle = \frac{1}{N^d}\frac{\partial \ln
Z_{\textrm{pq}}}{\partial \mu}  =  \frac{1}{N^d}\Big\langle
\frac{\partial_\mu C}{C}\Big \rangle_{\text{pq}}
$$
respectively. The average phase factor is
$$ \langle e^{i\varphi}\rangle_{\text{pq}} = \Big\langle \frac{C}{|C|}\Big\rangle_{\text{pq}} = \frac{\sum_{k=(k_{x,\alpha})}   U^{j}
C(x_1,\cdots,x_{2j})}{\sum_{k=(k_{x,\alpha})}
U^{j}|C(x_1,\cdots,x_{2j})|} $$ Here the average $\langle \
\rangle_{\text{pq}} $ is taken with respect to the partition
function $ Z_{\textrm{pq}}$ in (\ref{2016_7_2_5}).

\end{document}